\documentstyle[aps,prb,preprint]{revtex}
\begin{document}
\title{On the accuracy of the density matrix renormalization 
       group method}
\author{\"Ors Legeza}
\address{Research Institute for Solid State Physics, 
         H-1525 Budapest, P.\ O.\ Box 49, Hungary}
\address{Technical University of Budapest, 
         H-1521 Budapest, Hungary}
\author{G\'abor F\'ath\cite{byline}}
\address{Institute of Theoretical Physics, University of Lausanne,
         CH-1015 Lausanne, Switzerland}
\date{\today}
\maketitle
\begin{abstract} 
White's density	matrix renormalization group ({DMRG}) method has 
been applied to the one-dimensional Ising model in a transverse 
field ({ITF}), in order to study the accuracy of the numerical 
algorithm. Due to the exact solubility of the {ITF} for any finite 
chain length, the errors introduced by the basis
truncation procedure could have been directly analysed. By computing 
different properties, like the energies of the low-lying levels or the 
ground-state one- and two-point correlation functions, we obtained a 
detailed picture, how these errors behave as functions of the various 
model and algorithm parameters. Our experience with the {ITF} contributes 
to a better understanding of the {DMRG} method, and may facilitate its 
optimization in other applications.
\end{abstract}
\bigskip
\narrowtext
\section{Introduction}

In the last three years we have witnessed a breakthrough in the 
numerical analysis of one-dimensional (1D) quantum lattice models.
This considerable progress was due to the invention of the density
matrix renormalization group ({DMRG}) method by S.\ White.\cite{white} 
Recent applications of the method has demonstrated its extreme efficacity 
and versatility. By now the {DMRG} has become one of the leading numerical 
tools in the study of most 1D quantum spin and electron problems
of current interest.

There are several numerical methods to obtain the low-energy states
of a given quantum Hamiltonian. Exact diagonalization algorithms are
able to compute the ground state and the lowest excited states with a
precision of more than 12 digits. The attainable system sizes, however, 
are rather limited due to memory restrictions. On the other
hand, stochastic methods, like the diverse variants of the quantum
Monte Carlo method, are capable of treating systems with hundreds of
sites -- at the price of reducing the precision of the obtained
results. In this case, it is rather the computation time that limits
the applications.  Use of renormalization group procedures, such as
the {DMRG}, could be the way out of this dilemma. The key idea is to
gradually increase the system size, and, at the same time,
systematically truncate the Hilbert space by keeping only those
degrees of freedom that are really ``important'' for an accurate 
representation of the desired state. The main point, and this is
where the {DMRG} differs drastically from the preceding numerical RG
methods, is how to choose the most important degrees of freedom to
minimize the error caused by discarding the other ``unimportant''
ones.

The {DMRG} is an iterative algorithm to build up the lattice to the
desired length and find approximants to the target state (the
ground state or an excited state), using only a limited number of 
basis states. The total 
system is divided into two parts, the block and the environment. In
each step of the procedure the block is increased by adding one lattice
site, but only the most probable states, obtained from the reduced
density matrix of the block, are kept. The environment is composed 
from the block of the previous step, and its role is to embed
the block into a larger system, when the density matrix is formed. 
Application of the reduced density matrix in the truncation process,
and not simply keeping the lowest energy states of the block as in 
previous RG techniques,\cite{wilson} is the key ingredient of the 
{DMRG} method, since, as was shown by White, \cite{white} this is the 
way to minimize the error introduced into the representation of the 
target state. Although, during the algorithm the length of the total 
system increases gradually, the dimension of the Hilbert 
space is always kept manageable by the truncation process, 
and systems with relatively large size can be studied.

The {DMRG} has been successfully applied to various 1D and coupled
chain problems, such as $S=1/2$ \cite{white,s_half} 
and $S\ge 1$ \cite{s_one} spin chains, 
strongly correlated electron systems,\cite{fermi} impurity 
problems\cite{impur} or the two-chain Heisenberg 
and Hubbard models.\cite{2chain} 
Promising implementations of the algorithm to compute dynamical 
properties\cite{dyna} of 1D systems or to simulate 2D 
lattices\cite{2dima,2dimb} have also been reported. 
At least for quasi 1D problems,
its advantage over the standard numerical procedures has become
evident. While the available system sizes are comparable
to those of the Monte Carlo method, the precision of the computed
quantities seems better by several orders of magnitude. 

The {DMRG} works especially well when the system is subject to open
boundary condition. In the case of periodic boundary condition, on
the other hand, errors are definitely stronger. Moreover, the
conservation of momentum cannot be directly built into the method,
so restriction of the computation to a certain momentum sector is
not possible. These drawbacks has led to the fact that a large part of 
current applications of the method treat {\em open} systems. 

Although this is a renormalization group procedure, contrary
to the naive expectations, results seem more accurate if the system 
is away from criticality, i.e., when the model possesses a finite 
spectral gap in the thermodynamic limit. Other parameters of the 
model in study, like the number of degrees of freedom at a single 
site or the range of interactions also have drastic effects on the 
precision. 

For any numerical method, it is of essential importance to understand,
how errors evolve, when parameters of the procedure or those of the
model in study are varied. So far, the accuracy of the {DMRG} method has
only been tested on small clusters where numerically exact results
were known from exact diagonalization, or alternatively, large
lattice {DMRG} data were extrapolated to the infinite chain limit, and
these extrapolated values have been compared to rigorous results,
available only in the thermodynamic limit. In many of the
applications, however, at most the small lattice exact data are
available for testing. Nevertheless, large lattices, even with hundreds 
of sites, are used to draw conclusions on the behavior of the system. 
In many cases, it is really a difficult problem to reliably estimate 
the precision as the system size increases. 

Our main goal in this paper is to study the general trends of numerical 
errors during the {DMRG} algorithm, especially when the length of the
system is in the range of the typical applications, i.e., well beyond the
limit where exact diagonalization data are attainable.
For this purpose, we apply the method to an exactly solvable system, 
the {\em Ising model in a transverse field\/} ({ITF}). 
The main advantage of the {ITF} model as a test system
is that its energy spectrum and ground-state correlation
functions can be calculated by simple tools for {\em any finite}
chain length, even when the chain is subject to {\em open} boundary
condition, as in the standard {DMRG} applications.  The {ITF} is also a
model with a second order quantum phase transition, so the effect of
the appearance of criticality on the accuracy can also be analysed.

We carried out a detailed {DMRG} study of the {ITF}, varying several
parameters of the model and the numerical procedure, like the
strength of the magnetic field, the chain length, the number of 
states kept, the number of iterations and the number of target
states. Beside the energies of the low-lying states, we also computed
different one- and two-point correlation functions. 

The setup of the paper is as follows. In Sec.\ II we briefly summarize
the exact solution of the {ITF} with open boundary condition. Section
III is devoted to the details of the {DMRG} algorithm, while Sec.\ IV 
contains the analysis of the observed trends in the numerical errors.
Section V is a summary of our main conclusions.

\section{The 1D {ITF} with open boundary condition}

The Ising model in a transverse field ({ITF}) on a one-dimensional 
chain of $N$ sites is defined by the following Hamiltonian
\begin{equation}
{\cal H}= -\sum_{n=1}^{N-1} \sigma_n^x \/ \sigma_{n+1}^x 
          -\gamma \/ \sum_{n=1}^N \sigma_n^z,
\label{eq:ITF}
\end{equation}
where $\sigma^{\alpha}$, $\alpha=x,y,z$ are the Pauli matrices,
$\gamma$ is the transverse magnetic field applied in the $z$-direction,
and the chain is subject to {\em open} boundary condition, as in our
{DMRG} implementation.

In the thermodynamic limit, the {ITF} possesses a second order phase
transition that takes place at $\gamma=1$, where the energy gap vanishes
as $N\to\infty$. For $\gamma<1$ (ordered phase), the ground state is
doubly degenerate and there is a long-range order: the ground-state 
correlation function $\rho^x_l=\langle \sigma^x_n \sigma^x_{n+l}\rangle$
tends to a finite value as $l\to\infty$. On the other hand, for 
$\gamma>1$ (disordered phase), the ground state is unique and 
$\rho^x_l\to 0$ as $l\to\infty$. The gap is finite on both sides of the 
transition point.

The {ITF} in Eq.\ (\ref{eq:ITF}) is exactly solvable for any chain 
length $N$. Details of the calculation can be found, e.g., in 
Ref.\ \onlinecite{cab-jul}. Here we only summarize the main
steps and the necessary formulae. Introducing Fermi operators via
the Jordan-Wigner transformation 
\begin{eqnarray}
c_n&=&\exp \left( \pi\/i\sum_{j=1}^{n-1}\sigma_j^+\sigma_j^- \right) 
     \/\sigma_n^- \/, 
\end{eqnarray}
the Hamiltonian $\cal H$ reduces to a quadratic form
\begin{equation}
{\cal H}=-2\gamma\sum_{n=1}^N c_n^{\dag}c_n
     -\sum_{n=1}^{N-1}(c_n^{\dag}c_{n+1}^{}+c_{n}^{\dag}c_{n+1}^{\dag}
                       +h.c.) +N\gamma\,.
\label{eq:quadratic}
\end{equation}
This can be diagonalized directly by a Bogoliubov 
transformation\cite{lie-sch-mat} 
\begin{eqnarray}
\eta_k &=
& \/\sum_{n=1}^N \left( \frac{\phi_n^k+\psi_n^k}{2} c_n
+ \frac{\phi_n^k-\psi_n^k}{2} c_n^{\dag} \right) \,,
\label{eq:bogol}
\end{eqnarray}
leading to the final diagonal form
\begin{eqnarray}
{\cal H}\/&=&
\/ \sum_{k\,\mbox{\rm\scriptsize allowed}}  
   \Lambda(k) \left(\eta_k^{\dagger} \eta_k^{}- \frac{1}{2} \right),\\
\label{eq:diag}
\Lambda(k) &=&
2 \left( 1+\gamma^2+2\gamma\cos{k} \right)^{1/2}.
\label{eq:lambdareal}
\end{eqnarray}
The $N$-element real vector {\boldmath $\phi$}$^k$ 
is a solution of a set of linear 
equations,\cite{cab-jul} and turns out to be
\begin{eqnarray}
\phi_n^k &=&
A_N\left(\sin{kn}-\frac{\sin{k}}{\gamma+\cos{k}}\cos{kn}\right), 
\end{eqnarray}
where $n=1,2,\ldots,N$ and $A_N$ is a normalization constant. 
{\boldmath $\psi$}$^k$ can be expressed from {\boldmath $\phi$}$^k$ as
\begin{eqnarray}
\psi_n^k &=&
A^{\prime}_N\left(\gamma \phi_n^k +\phi_{n+1}^k \right), 
\label{eq:psi_nk}
\end{eqnarray}
where $A^{\prime}_N$ is another normalization constant, and 
formally $\psi_N^k\equiv 0$.

Reflecting the fact that the chain is subject to open boundary condition,
the allowed $k$ modes in the summation of Eq.\ (\ref{eq:diag}) cannot be
written in a simple form for general $\gamma$, but are
determined through a trigonometric equation 
\begin{equation}
{\frac{\sin[k(N+1)]}{\sin(kN)}}=-\frac{1}{\gamma}.
\label{eq:realk}
\end{equation}
In any case, the total 
number of independent modes is equal to the number of sites $N$. 
By convention, the allowed k values are the roots $k=k_0+ik_1$, whose
real part $k_0$ is in the interval $0<k_0\le\pi$. 
When $\gamma>\gamma_c(N)\equiv N/(N+1)$ all the roots are real.
For $\gamma<\gamma_c(N)$, however, a complex ({\em localized\/}) solution
becomes possible with $k_0=\pi$ and $k_1>0$. Note that $\gamma_c(N)\to 
\gamma_c=1$, the phase transition point, as $N\to\infty$. The  
imaginary part $k_1$ can be obtained by solving the equation
\begin{equation}
{\frac{\sinh[k_1(N+1)]}{\sinh(k_1N)}}=\frac{1}{\gamma},
\label{eq:complexk}
\end{equation}
and the corresponding {\boldmath $\phi$} 
vector is \cite{cab-jul}
\begin{eqnarray}
\phi_{n}^{\mbox{\scriptsize loc}} &=&
A_N (-1)^{n-1} \sinh \left[ (N-n+1)k_1 \right], 
\end{eqnarray}
where, again, $n=1,2,\ldots,N$ and $A_N$ is a normalization constant.
$\psi_n^{\mbox{\scriptsize loc}}$ is obtained again through Eq.\ 
(\ref{eq:psi_nk}).
The elements of $\phi_n^{\mbox{\scriptsize loc}}$ and 
$\psi_n^{\mbox{\scriptsize loc}}$ are
exponentially small far from the left and right chain ends, respectively,
which indicates, by Eqs.\ (\ref{eq:bogol}), that the complex mode is 
localized to the chain ends.
The energy associated with this localized mode is 
\begin{equation}
\Lambda(\pi+ik_1)=2 \left( 1+\gamma^2-2\gamma\cosh{k_1} \right)^{1/2}.
\label{eq:lambdacomplex}
\end{equation}
When $N$ is large, Eq.\ (\ref{eq:complexk}) is readily solved to
yield  $k_1\sim \ln 1/\gamma$. By substitution, this gives 
$\Lambda(\pi+ik_1)\sim 0$, i.e., it is a  
mode of zero energy, leading to the double degeneracy of the 
ground state sector when $\gamma<1$. 

The ground state energy of the model, the fermionic vacuum, is 
expressed as
\begin{equation}
E_0=\frac{1}{2} \sum_{k\,\mbox{\rm\scriptsize allowed}} \Lambda(k),
\label{eq:gsenergy}
\end{equation}
that can be written in a closed form only at the transition point 
$\gamma=1$, where it reduces\cite{bur-gui} to
\begin{equation}
E_0=1-\mbox{cosec} \left(\frac{\pi}{4N+2} \right).
\end{equation}
The first excited state of the model is always determined by the wave 
number $k$, whose real part $k_0$ is closest to $\pi$; for 
$\gamma>\gamma_c$ it is a real root
and the energy gap converges to $2(\gamma-1)$ as $N\to\infty$. For 
$\gamma<\gamma_c$, this is the localized mode, and the first excited state
becomes asymptotically degenerate with the ground state. It is the second
excited state that constitute the real (finite) energy gap in this case.  

Finally, ground-state one- and two-point correlation functions can also be
calculated, by the method described in details by Lieb {\em et al.\/} 
\cite{lie-sch-mat}. The only interesting one-point function is 
${\cal M}_n^z=\langle \sigma^z_n\rangle$, which can be expressed as
\begin{equation}
{\cal M}_n^z=\frac{1}{2}G_{nn},
\label{eq:Miz}
\end{equation}
where $G_{nm}$ is defined by
\begin{equation}
G_{nm}=-\sum_{k\,\mbox{\rm\scriptsize allowed}} \psi_n^k\phi_m^k.
\label{eq:G}
\end{equation}
Note that even in the ordered state 
${\cal M}_n^x=\langle \sigma^x_n\rangle=0$,
since the Hamiltonian, and thus any finite lattice ground state, is 
invariant to the transformation $\sigma_n^x\to -\sigma_n^x$. The same
holds for $\sigma_n^y$, so that ${\cal M}_n^y=\langle \sigma^y_n\rangle=0$, 
too. As for the two-point functions, 
$\rho_{nm}^x=\langle \sigma^x_n \sigma^x_m\rangle$,
$\rho_{nm}^y=\langle \sigma^y_n \sigma^y_m\rangle$ and
$\rho_{nm}^z=\langle \sigma^z_n \sigma^z_m\rangle$, one arrives to the
formulae
\begin{eqnarray}
\rho_{nm}^x&=& 1/4
\left|\matrix{ G_{n,n+1} & G_{n,n+2} & \cdots & G_{nm} \cr
                         \vdots & & & \vdots \cr
               G_{m-1,n+1} &           \cdots & & G_{m-1,m} \cr } \right|,
\label{eq:ro_x} \\
\rho_{nm}^y&=& 1/4
\left|\matrix{ G_{n+1,n} & G_{n+1,n+1} & \cdots & G_{n+1,m-1} \cr
                         \vdots & & & \vdots \cr
               G_{m,n} &               \cdots & & G_{m,m-1} \cr } \right|,
\label{eq:ro_y} 
\end{eqnarray}
and
\begin{equation}
\FL
\rho_{nm}^z = 1/4
\left(G_{nn}G_{mm}-G_{mn}G_{nm}\right) \,,
\label{eq:ro_z}
\end{equation}
where $n<m$.
 
In order to be able to compare the approximate results, produced by the
{DMRG} method, with the exact ones for a certain value of $\gamma$ and
chain length $N$, the exact quantities were calculated by an independent 
numerical process, using either double precision {\em Fortran} routines or
the {\em Mathematica} software that allows computations with arbitrary
precision. We solved the nonlinear equations (\ref{eq:realk}) and 
(\ref{eq:complexk}) to obtain the allowed set of wavenumbers, and used
these values to express the associated {\boldmath $\phi$}$^k$ and 
{\boldmath $\psi$}$^k$ vectors. Then the ground-state and excited 
state energies and the correlation functions were calculated numerically 
according to the above formulae. All the results obtained in this 
way were precise to at least 14 digits.

\section{The Density-Matrix Renormalization Group method}
 
White's density matrix renormalization group meth\-od (DMRG) 
is a numerical real-space renormalization group procedure, in 
which the effective size of the system increases gradually, 
while the dimension of the associated Hilbert
space remains constant, due to a systematic truncation process. 
Since the method is well described in the original papers,\cite{white}
we only present a brief summary here.

The basic object of the method is the system block $B_l$ that consists 
of $l$ lattice sites. All the necessary operators ${\cal A}_i$ of this 
block (e.g., $\sigma^{\alpha}_n$, $\alpha=x,y,z$ and $1\le n\le l$) are 
stored as $M\times M$ matrices. Note that the real dimension of a block 
of $l$ sites is $d^l$, where $d$ is the number of states at a single 
site, and $M<<d^l$, so the representation of the block matrices is
only approximative. In each step of the algorithm, a new single site 
$\bullet$ is added to the existing block $B_l$, and operators of
the resulting system $B_l\bullet$ of dimension $Md$ are formed as tensor 
products from the matrix representations of the corresponding operators 
of the two constituting parts. $B_l\bullet$ is then renormalized, first 
by carrying out an appropriate unitary transformation, then by truncating 
its degrees of freedom from $Md$ to $M$. Only the most important 
$M$ states are kept, and the remaining $M(d-1)$ ones are discarded.
The resulting system of dimension $M$ and length $l+1$, denoted by
$B_{l+1}$, is then used iteratively in the subsequent step of the algorithm.
Choosing the states kept, i.e., finding the appropriate unitary 
transformation, is the most crucial point of the renormalization 
group procedure. 

Previous applications of the renormalization group technique, where the $M$
lowest energy states of $B_l\bullet$ were kept, led to disappointing
results for many model systems. As was observed by several 
authors,\cite{igl-whi-noa} this was due the interaction having been 
neglected during the renormalization process between $B_l\bullet$ and 
the rest of the larger (in many cases infinite) system of what it makes a 
part. $B_l\bullet$ was renormalized in a way that an unnatural 
open boundary condition was forced on both of its ends. 

In order to avoid this problem,
White's method embeds $B_l\bullet$ into a superblock, and uses the 
eigenstates of this larger system to carry out the renormalization of
$B_l\bullet$. Choosing the superblock configuration as $B_l\bullet\bullet 
B_l^R$, where $B_l^R$ is the reflection of $B_l$, its effective 
Hamiltonian, built up from the matrix representations of the necessary
operators of the two blocks and the two single sites, is diagonalized to 
obtain the desired (target) state $\Psi$. Even though, the target state 
can be a linear combination of more states, by targeting only one state,
the renormalized block states are more specialized for representing 
that single one, and fewer of them are needed for a given accuracy.

Having the target state expressed on the superblock as
\begin{equation}
\Psi = \sum_{i,j=1}^{Md} \Psi_{i,j} |i\rangle |j\rangle ,
\end{equation}
where $i$ and $j$ label the $Md$ states of $B_l\bullet$ and its 
surroundings $\bullet B_l^R$, respectively, the reduced density matrix
of the subsystem $B_l\bullet$ is formed as
\begin{equation}
\rho_{ii'}=\sum_j\Psi_{ij}\Psi_{i'j}.
\end{equation}
As was shown by White,\cite{white} the error introduced 
into the representation of
the target state by the truncation to $M$ states is minimized, if the 
new basis that one changes to is the basis that diagonalizes the
density matrix $\rho_{ii'}$. The renormalization of the tensor product
operators of $B_l\bullet$,
\begin{equation}
{\cal A}_i \to {\cal O A}_i {\cal O}^{\dagger},
\end{equation}
is thus carried out by the $M\times Md$ 
transformation matrix $\cal O$, whose rows are composed from the $M$
density matrix eigenvectors, associated to the $M$ largest 
eigenvalues $\omega_{\alpha}$, $\alpha=1,\cdots,M$. 

The simplest version of the {DMRG} method, the {\em infinite lattice 
algorithm}, starts with four lattice sites, i.e., from the
superblock configuration $B_1\bullet\bullet B_1^R$. In each step,
the total length of the chain increases by 2. Measurements of the 
interested quantities are made after the calculation of the target 
state in each step, and the whole process is continued until the
results converge satisfactorily. This method is especially suitable
to yield, with minimal computational efforts, rather precise estimates
of bulk properties, like the ground-state energy density or ground-state
correlation functions.

The usual measure of the 
accuracy of the truncation to $M$ states is the deviation of the
sum of the density matrix eigenvalues associated with the states kept
$P_M \equiv \sum_{\alpha=1}^M \omega_{\alpha}$ from unity. Clearly,
in the extreme case when the discarded states have zero weight, i.e.,
$\omega_{M+1}=\omega_{M+2}=\cdots=\omega_{Md}=0$ so that $P_M=1$, they
are not required to represent the target state $\Psi$,
and no error has been committed.
There is, however, another source of error in the {DMRG}
procedure, namely that the $B_l\bullet$ subsystem is only embedded into
an {\em approximate} superblock when it is renormalized, and not into
the, a priori unknown, exact environment. The two kinds of errors,
the "truncation" error and the "environment" error are not simply 
additive. The latter one, however, can be reduced in an iterative 
manner, using the so called {\em finite lattice algorithm\/}.

The finite lattice algorithm starts by 
building up the lattice to the desired length N, using the infinite 
lattice algorithm. Then the superblock configuration is modified to
$B_l\bullet\bullet B_{N-2-l}^R$, so that the total length remains 
always $N$, and the blocks $B_l$, $l=1,\cdots, N-3$, are recomputed. 
The process when all $B_l$'s are recomputed constitute the {\em cycle}
of the algorithm.
Since at each step from this time on, the block to be renormalized 
is part of a system of the desired length $N$, the environment becomes
more precise, and a considerable improvement of
the results, corresponding to the $N$-site system, is achieved. These 
results can be then used to carry out a systematic finite-size scaling
analysis, or to study, e.g., boundary effects in the finite chain.

\section{Numerical results}

In order to determine the accuracy of {DMRG} method, numerical 
calculations on the {ITF}, using both the infinite and finite 
lattice algorithms, were performed. Errors in various quantities, 
such as the ground-state and first excited-state energies
$E_{\rm GS}$ and $E_{\rm 1XS}$, resp.,
the one-point correlation functions ${\cal M}^{z}_{n}$ and 
${\cal M}^{x}_{n}$, and the two-point correlation functions 
$\rho^{z}_{l}$ and $\rho^{x}_{l}$ were monitored. In the case 
of the {\em infinite} system algorithm, our main concern was
how the errors of the energies depend on the system size $N$, 
when other parameters are kept constant. In the {\em finite} lattice 
algorithm, on the other hand, the length of the 
chain was kept fixed, and the effect of introducing additional cycles in
the process was analyzed. When excited states were computed, 
the number of target states was also varied. 

Since the {ITF} possesses a second order phase transition at 
$\gamma_c=1$, it was expected that many of the above quantities have 
different behavior, whether $\gamma$ is equal to the critical value, or 
greater or less than $\gamma_{c}$. 
Therefore, we investigated the three different regions of the 
phase diagram by choosing three different values
for $\gamma$, namely, $\gamma=0.5<\gamma_c$, $\gamma=1=\gamma_c$, and
$\gamma=2>\gamma_c$.

\subsection{Energies}

Let us first consider the error of the ground-state (GS) and first 
excited-state (1XS) energies. Using the {\em infinite lattice algorithm}, 
we built up a chain of length $N=300$ and kept states up to $M=48$.
Except one example which will be discussed below, we found that
targeting one state alone is by orders of magnitude more precise 
that targeting several states together. Hence, unless stated otherwise, 
the results to be presented
here were obtained in a way that the ground state and the first 
excited state were targeted {\em separately}.

The errors $E(\mbox{\rm\scriptsize DMRG})-E({\rm exact})$ were always 
found to be positive, satisfying
the statement that the {DMRG} is a variational method, which gives 
(at least for the GS energy) an upper bound estimate.\cite{2dima} 
The relative errors 
$\delta E\equiv [E(\mbox{\rm\scriptsize DMRG})-E({\rm exact})]
/ \left| E({\rm exact}) \right|$ as a function of $N$ for various 
values of $M$ and $\gamma$ are shown on log-log scale in 
Fig.\ \ref{fig:en2}--\ref{fig:en1}. 

Until the algorithm keeps all states, i.e., for $N\le 2(\ln M+1)$, 
the {DMRG} is exact and only the machine's numerical inaccuracy is seen,
limiting the precision on the 
order of $10^{-14}$. For longer chains, however, the errors associated 
with the reduction of degrees of freedom also come in. 

For the off-critical values, $\gamma=2$ and $0.5$, the ground state 
error $\delta E_{\rm GS}$ shows practically no size dependence. 
It depends, however, crucially on the
value of $M$. The behavior is the most clear for $\gamma=2$, where the
infinite system ground state is unique with a gap above it. Even a small 
value of $M$, $M\sim 10$, is enough to reach the machine's precision 
limit [Fig.\ \ref{fig:en2}(a)]. There is, on the other hand, an 
interesting size dependence in the error of the first excited state 
$\delta E_{\rm 1XS}$, especially for greater $M$ values 
[Fig.\ \ref{fig:en2}(b)]. A maximum evolves, above which 
$\delta E_{\rm 1XS}$ begins to dwindle again. This can be understood, 
however, since the approximate excited state, yielded by the {DMRG} 
process, is not necessarily exactly orthogonal to the real ground state, 
and can have a finite overlap with it. This fact leads to an 
overcompensation of the otherwise increasing positive absolute error 
for long chains.

For $\gamma=0.5$, where the ground state is asymptotically doubly 
degenerate, the behavior is more subtle [Fig.\ \ref{fig:en5}(a,b)]. 
The splitting of the two lowest levels
decreases exponentially as $N$ increases. When this difference goes below
the machine's precision limit $\sim$$10^{-14}$, the algorithm is incapable
to further resolve the two levels, and the target state that it yields is
a linear combination of the two exact eigenvectors with random coefficients.
As a consequence, in each step of the {DMRG} 
algorithm the target state changes unpredictably, leading to the loss 
of optimality of the truncation process, and hence, to considerably higher, 
rather scattered error rates. As Fig.\ \ref{fig:en5}(b) shows, the 
{DMRG} can even "lose" the target state above a certain chain length.
In a try to avoid these problems, we also computed $E_{1XS}$ by 
choosing the target state to be a linear combination of the 
ground-state and the first excited-state vectors. 
This proved to be an improvement in the range where otherwise 
the excited state was lost; for smaller values of $N$, however, 
the errors were considerably larger [Fig.\ \ref{fig:en5}(b)]. 

At the critical point $\gamma=1$, the errors were found to be by 
several orders of magnitude larger than at the off-critical values of 
$\gamma$. This is in accordance with the findings of Ref.\ 
\onlinecite{drz-vanlee}: the correlation length of the model is one of
the most significant factors that influence the precision of the {DMRG}
method. Moreover, the curves in Fig.\ \ref{fig:en1}(a,b) have a clear
size dependence. When $M$ is kept fixed, the errors can become 
larger by 4 to 6 orders of magnitude, as the chain length approaches 
$N=300$. For smaller $M$, the clear downward curvature seen
in the log-log plot indicates that the errors converge to a finite 
value at large lengths. For larger $M$ values, however, this convergence
is much slower, and the analysis is made more difficult by the
appearance of a crossover effect, which change the behavior in the small
$N$ region, and makes the curves more flat there. While for $M=16$ the 
crossover size is around $N\sim 10$ (and hence unobservable), for $M=32$ 
it is at $N\sim 100$, and for $M=48$ it is at $N\sim 250$, showing that
the crossover size scales for larger lengths as more and more states are 
kept.

A possible interpretation of this crossover effect can be obtained by 
recalling that there are two different sources of errors in the {DMRG} 
method (see Sec.\ III). For small $M$, this is clearly the "truncation"
error that dominates. Curves with $M=4,8,16$ in Fig.\ 
\ref{fig:en1}(a,b) basically show the size dependence of this type of
error alone. The effect of the "environment" error only shows up for
large enough $M$ and small enough $N$ values, when the truncation error 
is strongly reduced. The environment error approaches its saturation 
earlier than the truncation error, as is seen from the $M=48$ curves, 
where the environment error dominates. However, since the truncation 
error increases several orders of magnitude in the
small $N$ regime, it always exceeds the environment error for large 
enough sizes. This produces a rather sharp break in the curves in the 
log-log plot at the crossover size, such as seen for $M=32$ in the figure.

For comparison, and to reduce the environment error, 
computations using the {\em finite lattice algorithm}
were also performed. Fixing the chain length at      
$N=100$, the iteration process was repeated until the desired energies 
converged. For $\gamma=2$, the relative errors are plotted in Fig.\
\ref{fig:pm2}(a,b) as a function of the sum of the discarded density 
matrix eigenvalues $1-P_m$. While there is practically no improvement
in the ground-state energy, the first excited-state energy becomes 
much more precise, when further cycles are carried out, and full 
convergence is reached only after the $I=3$ iteration. At the critical 
point $\gamma=1$, $\delta E_{\rm GS}$ and $\delta E_{\rm 1XS}$ behave
similarly: two cycles are needed to get rid of the environment error
[Fig.\ \ref{fig:pm1}(a,b)]. 

It is seen in the figures that the $I=1$ cycle data (the infinite 
lattice algorithm results) do not fit onto a straight line on the log-log 
plot. Points from the fully converged cycles, however, do so nicely. 
The slope of the fitted line was found to be very close to unity in 
all cases, indicating that the error is proportional to the 
discarded weights, i.e., 
\begin{equation}
\delta E={\rm const}\cdot (1-P_m), 
\end{equation}
where the constant can depend on the model parameters and the system size. 
We emphasize, however, that this form only holds for the 
{\em converged} energies. Extrapolating the infinite lattice algorithm 
($I=1$) data by this formula to the $1-P_m\to 0$ ($M\to\infty$) limit can
yield false results.

\subsection{One-point correlation functions}

Expectation values of local operators in the ground state were computed
using the finite lattice 
algorithm at a fixed chain length $N=100$. Both ${\cal M}^x_n$ and 
${\cal M}^z_n$ (n=1, \dots,100) were measured, and the errors 
$\delta {\cal M}^{\alpha}_n\equiv  
{\cal M}^{\alpha}_n(\mbox{\rm\scriptsize DMRG})
-{\cal M}^{\alpha}_n({\rm exact}) $, $\alpha=x,z$
were calculated. Note that for ${\cal M}^x_n$, the exact values
are zero, as it was detailed in Sec.\ II.

At the critical point $\gamma=1$, our results are presented in Fig.\
\ref{fig:mag}(a,b). The $I=1$ cycle (infinite lattice
algorithm) produces an error which depends significantly on the 
position in the chain $n$. Results are more precise 
around the middle of the system. Note that spatial variations 
can reach several orders of magnitude for larger $M$ values, 
as e.g.\ for the $I=1$, $M=16$ curve in Fig.\ \ref{fig:mag}(a). There 
is also a sudden improvement in accuracy very near to the chain ends, 
but this is believed to be an anomaly of the {ITF} model and not a general 
feature of the {DMRG} technique. (The exact one-point functions of the 
model, when open boundary condition is used, show a strong boundary
effect close to the ends, and this seems to influence the errors too.)

In the case of ${\cal M}^z_n$ [Fig.\ \ref{fig:mag}(a)], additional 
iteration cycles ($I\ge 2$) 
considerably improve the precision by decreasing the environment error.
Although some fluctuations may still be
present (especially for large $M$) in the $I=2$ data, the error 
becomes more or less constant and spatially more homogeneous for $I\ge 2$.
The convergence found with respect to the number of cycles is similar 
to that of the energies.

For ${\cal M}^x_n$ [Fig.\ \ref{fig:mag}(b)], where the exact values of
the magnetization are zero, the situation is different. Additional cycles,
and increasing the value of $M$, rather unexpectedly, 
make the data less precise.  The sign of the error also changes: 
while for the $I=1$ cycle $\delta {\cal M}^x_n$ is negative,
for $I=3$ it always turns out to be positive. (Note that in the figure
the absolute values of the errors are plotted.) The rapid oscillations 
seen in the $I=2$, $M=8$ curve are due to the fact that the error 
fluctuates between positive and negative values. 
We believe that for this pathological case the observed errors stem
from the numerical inaccuracy of the diagonalization subroutines, and not 
from the algorithm itself. Test runs on short systems keeping 
all states in the blocks, i.e., when the DMRG is numerically exact, 
produced the same qualitative picture of $\delta {\cal M}^x_n$.

It might have been thought that monitoring the magnitude of errors of
quantities, whose exact results are a priori known, like ${\cal M}^x_n$
in our case, could yield information on the precision on other, a priori
not known quantities, like ${\cal M}^z_n$. We see, however, that the 
errors of the two one-point functions behave completely differently
without any obvious correlation, so the knowledge of the accuracy of one
of them cannot be used to draw predictions on the other one.

For the $\gamma=2$ case (figure not presented), all the iteration 
cycles gives the same result for ${\cal M}^z_n$. There is practically
no $n$-dependence, and the
algorithm reaches the border of numerical inaccuracy very soon,
in accordance with what was found for the ground-state energy.
For $M=8$, the relative errors scatter on the scale of $10^{-9}$,
and this precision could not be improved by increasing $M$.
The behavior was found to be very similar in the ordered phase, 
at $\gamma=.5$, the only difference is that more cycles were needed 
for the full convergence.

\subsection{Two point correlation functions}

Measurements of the two-point correlation functions were carried
out similarly to the one-point correlation functions, at a fixed length
$N=100$. Following White's recipe, 
$\rho_l^{\alpha}\equiv \rho_{n,n+l}^{\alpha}$, $\alpha=x,z$, was 
measured so that the points $n$ and $n+l$ were positioned
symmetrically to the middle of the chain.\cite{white} This special
allocation of the points assures that, at least for the short-range 
correlations, end effects are strongly reduced. Although, in most
applications the bulk correlation functions are of interest,  since our 
aim was to test the {DMRG} algorithm itself, we compared the numerical 
data with the exact finite-lattice results obtained at $N=100$.

At the critical point, $\gamma=1$, the absolute value of the error of 
the correlation functions $|\delta \rho^{\alpha}_l| \equiv 
|\rho_l^{\alpha}(\mbox{\rm\scriptsize DMRG})
-\rho_l^{\alpha}({\rm exact})|$ is plotted in Fig.\ \ref{fig:cor}(a,b) 
for $\alpha=z$ and $x$. For small $M$ values, the curves are rather 
smooth with a moderate dependence on $l$. Convergence with respect 
to the number of cycles $I$ is reached when the ground-state energy 
converges. For larger values of $M$, the errors seem to fluctuate 
rather irregularly, especially for $|\delta \rho^x_l|$, for which 
several cusps evolve, where the error rate drops abruptly. This strange 
feature, however, is once again an artifact of plotting the {\em 
absolute value} of the error on a logarithmic scale. For each $l$ where 
the cusp appears, we found that $\delta \rho^{\alpha}_l$ changes sign, 
similarly to the behavior of the corresponding ${\cal M}^x_n$ curve, 
and this sign fluctuation causes the strange-looking 
shape of the curves.

Disregarding the fact that the sign of the errors is not fixed, there is
a clear tendency that in the $I=1$ cycle (infinite lattice 
algorithm) the long-range correlators are less precise. This is in 
complete accordance with the behavior of the one-point functions: 
operators are represented less accurately moving towards the chain ends 
because of the environment error. More cycles improve the situation,
especially for $\delta \rho^z_l$ [Fig.\ \ref{fig:cor}(a)], 
where the errors for large $l$ decrease by
3 orders of magnitude and, quite unexpectedly, the long-range 
correlators become more precise than the short-range ones. On the other
hand, there is no similar change in the tendency of the curves for 
$\delta \rho^x_l$ [Fig.\ \ref{fig:cor}(b)]. Although the improvement is
the most significant for large $l$, long-range correlators remain less
accurate.

For the noncritical values $\gamma=2$ and $.5$ (figure not presented), 
$\delta \rho^z_l$ and
$\delta \rho^x_l$ show practically no $l$-dependence. There is
only a moderate change in accuracy for very small and very large
values of $l$, but this does not exceed an order of magnitude.
The average accuracy of the correlation functions is considerably
worse than that of the energies. The limit of the relative precision
we could achieve by increasing $M$ was not better that $10^{-9}$,
similarly to what was found in the case of the one-point functions.
The effect of carrying out more cycles was also similar, the correlation
functions converged exactly when the energies reached their limit
values.

\section{Summary}

In the present paper, we have analysed in detail the accuracy of 
White's density matrix renormalization group method,
by applying it to the one-dimensional Ising model in a 
transverse field. Due to the exact solvability of this model,
the exact and the numerical results could have been directly compared. 
We varied several parameters, either in the numerical algorithm 
or in the model, and obtained a rather detailed picture how the accuracy
of the {DMRG} approximation depends on these parameters.

Our main results are summarized as follows:

(i) The {DMRG} yields an extremely precise value for the ground-state 
energy, especially when the model is far from the critical point, 
and the ground state is unique. Excited state energies or the ground 
state energy of a critical system can be obtained with much less accuracy. 

(ii) Targeting exclusively one of the excited states is, in general,
more precise than targeting it together with the ground states or other
low-energy states. This, however, can become unstable when the levels 
are too close to each other, like in the case of asymptotic degeneration
of the ground state. When this is expected to happen, more states must be
targeted together.

(iii) Carrying out only the $I=1$ iteration cycle (infinite lattice 
algorithm) leads to a significant "environment" error. Its effect is 
the most pronounced, when lots of states 
are kept, i.e., when $M$ is large, so the "truncation" error is relatively 
small. The two types of error produce a crossover effect as $N$ increases
in the critical case.
The environment error dominates in the small $N$ large $M$ regime.

(iv) Although the chain length dependence of the errors is strong in the
critical case, both the truncation and the environment errors seem to
converge to finite values as $N$ increases. In the practical range of 
applications $10<N<300$, however, the errors, especially the truncation
error, become larger by 2--3 orders of magnitude.

(v) For large $M$ values, the {\em finite} lattice algorithm ($I\ge 2$)
improves the results considerably by decreasing the environment error. 
This is seen not only in the energies, but in the one- and
two-point correlation functions. When the data are fully converged 
(but not after the first $I=1$ cycle), the 
errors are nicely proportional to the sum of the discarded density 
matrix eigenvalues $1-P_m$. This fact can be used to make an extrapolation 
to the $M\to\infty$ limit.

(vi) Accuracy of the correlation functions is always worse than that of 
the energies. In the $I=1$ cycle, the error in the representation of the 
local operators become larger as one moves outward from the chain center.
When two-point functions are computed in the usual way, i.e., 
symmetrically to the chain center, this leads to the 
fact that the long-range correlators are less precise. Additional 
iteration cycles make the errors smaller and their spatial dependence 
more homogeneous.

Although the one-dimensional {ITF} is a rather simple many body system, we
believe that most of our findings hold equally well for other, more complex
quasi-one-dimensional lattice problems. We hope that the above results 
contribute to a better understanding of the {DMRG} procedure, and provide
a direct help in optimizing the algorithm in other applications.

\section{Acknowledgments}

The authors would like to thank Jen\H o S\'olyom for encouragement,
helpful discussions and the critical reading of the manuscript.
This research was supported in part by the Hungarian Research Fund
(OTKA) Grant Nos.\ 2979, 15870, and by the Swiss National Science 
Foundation Grant No.\ 20-37642.93. \"OL gratefully acknowledges the 
hospitality of the Institute of Theoretical Physics of the University 
of Lausanne, where this work was started.

\newpage

\begin{figure}
\caption{Relative errors of the (a) ground-state energy, and (b) first
excited-state energy as a function of the chain length $N$ for
different values of $M$. Data obtained by the infinite lattice
algorithm ($I=1$) in the disordered regime at $\gamma=2$.} 
\label{fig:en2} 
\end{figure}

\begin{figure}
\caption{Same as Fig.\ \ref{fig:en2}, but in the ordered regime 
at $\gamma=0.5$. Curves labelled by ${\rm TS}=2$ show the case, 
when the excited state was targeted together with the ground state.} 
\label{fig:en5} 
\end{figure}

\begin{figure}
\caption{Same as Fig.\ \ref{fig:en2}, but at the critical point 
$\gamma=1$. Curves with $M=4,8,16$ are dominated by the truncation 
error, while those with $M=48$ by the environment error. The $M=32$
curves show a crossover between the two types.} 
\label{fig:en1} 
\end{figure}

\begin{figure}
\caption{Relative errors of (a) the ground-state energy, and (b) first
excited-state energy as a function of the sum of the discarded density
matrix eigenvalues $1-P_m$ for different values of $M$ (shown as labels). 
Data obtained by the finite lattice algorithm, carrying out the $I=1,2,3$ 
iteration cycles, and the chain length is fixed at $N=100$. 
Disordered regime at $\gamma=2$.} 
\label{fig:pm2} 
\end{figure}

\begin{figure}
\caption{Same as Fig.\ \ref{fig:pm2}, but at the critical point 
$\gamma=1$.} 
\label{fig:pm1} 
\end{figure}

\begin{figure}
\caption{Errors of the one-point functions (a) ${\cal M}^z_n$, and (b) 
${\cal M}^x_n$ as a function of the position in the chain $n$.
Data obtained by the finite lattice algorithm, carrying out the $I=1,2,3$ 
iteration cycles, and keeping $M=8$ or $16$ states. 
The chain length is fixed at $N=100$. Critical point $\gamma=1$.} 
\label{fig:mag} 
\end{figure}

\begin{figure}
\caption{Errors of the two-point functions (a) $\rho^z_l$, and (b) 
$\rho^x_l$ as a function of $l$.
Data obtained by the finite lattice algorithm, carrying out the $I=1,2,3$ 
iteration cycles, and keeping $M=8$ or $16$ states. 
The chain length is fixed at $N=100$. Critical point $\gamma=1$.} 
\label{fig:cor} 
\end{figure}

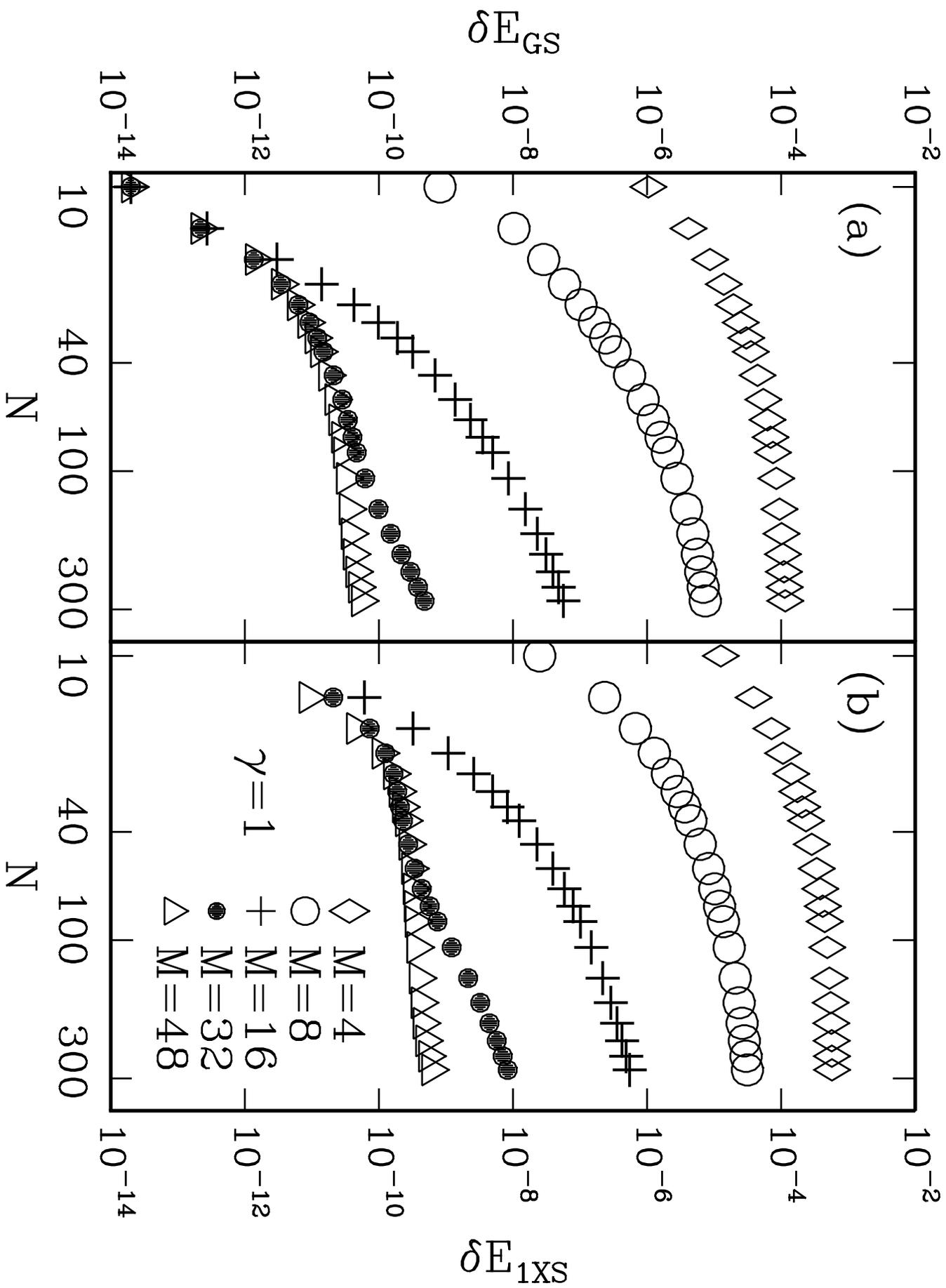
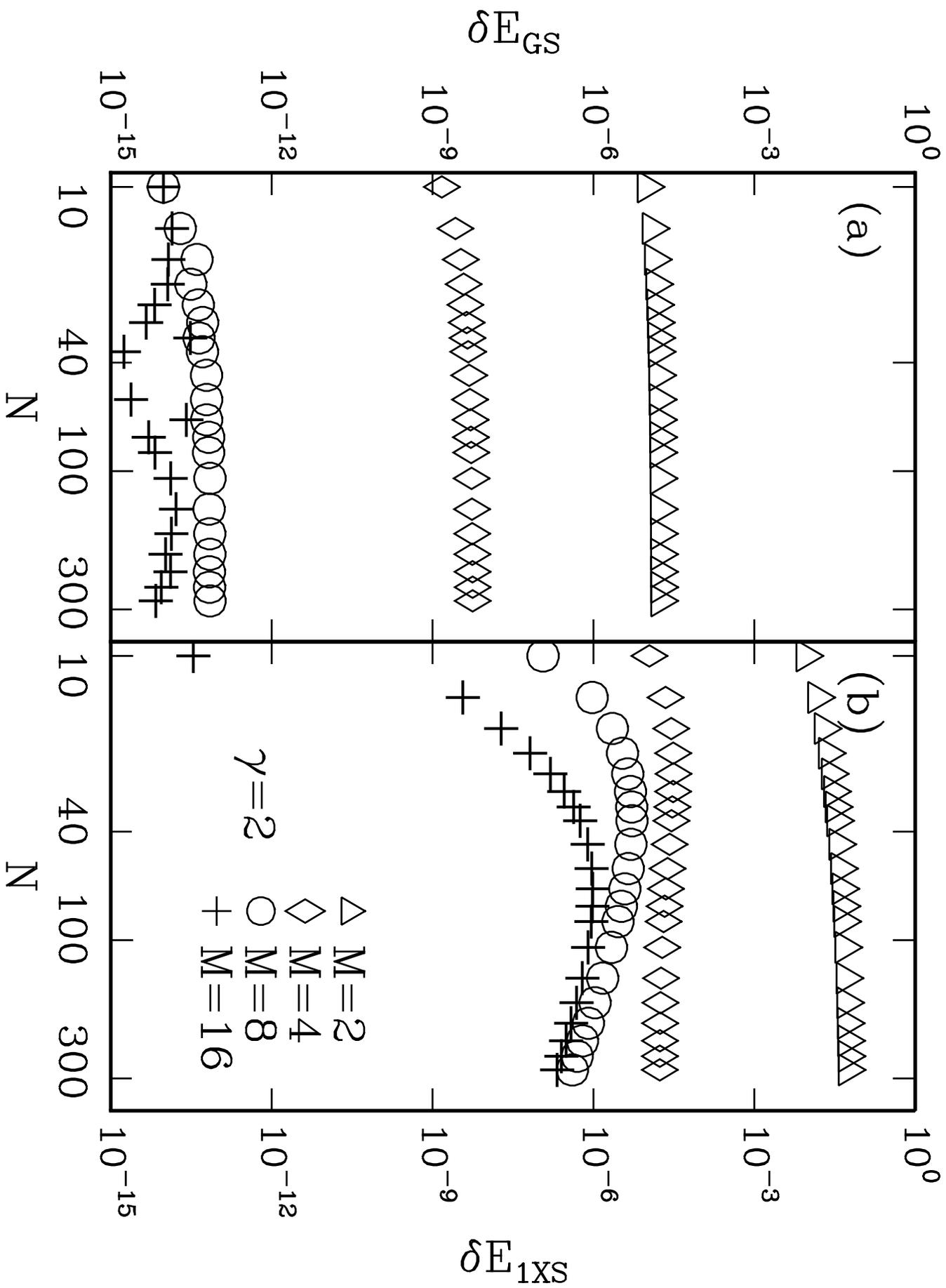
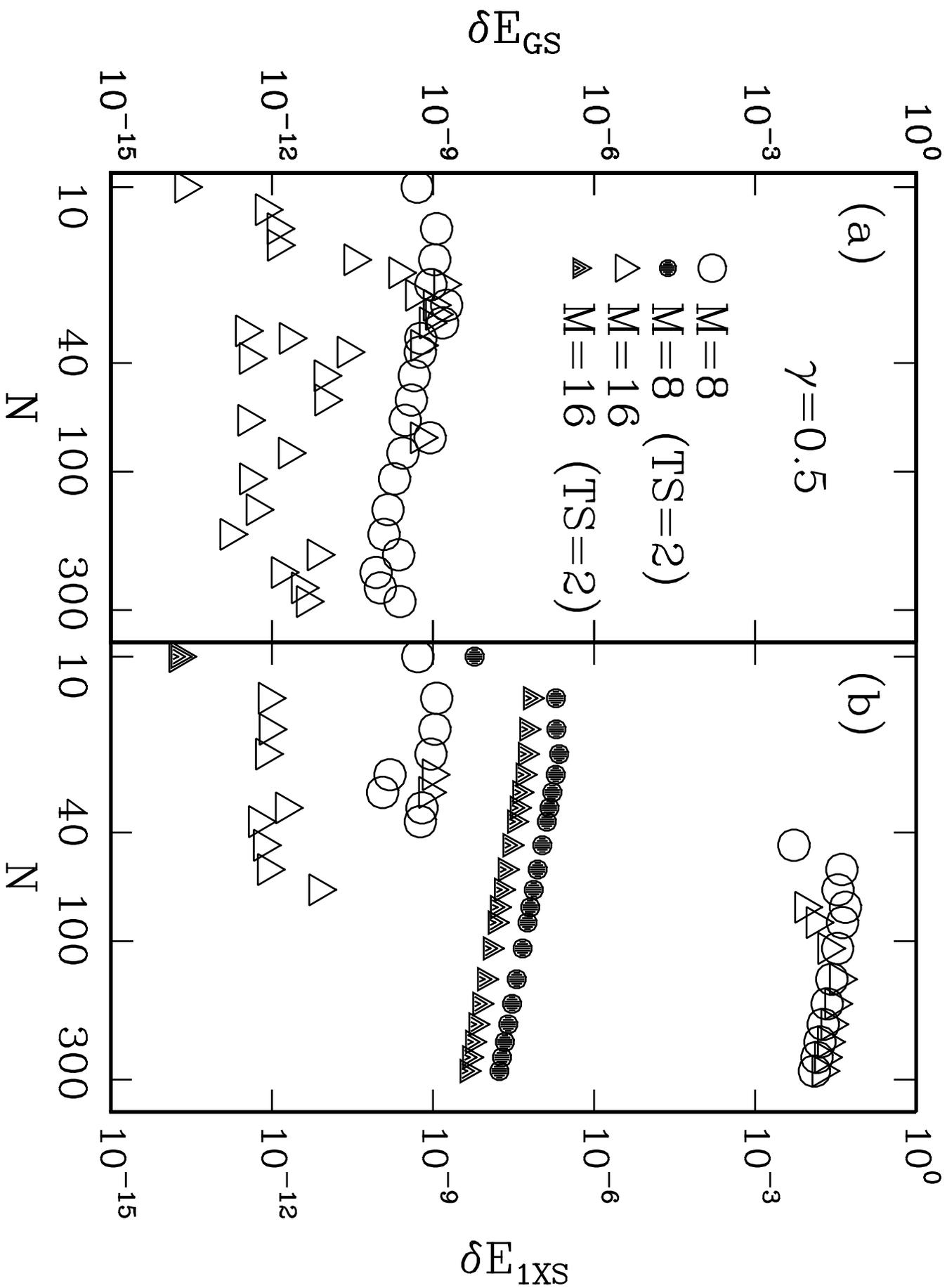
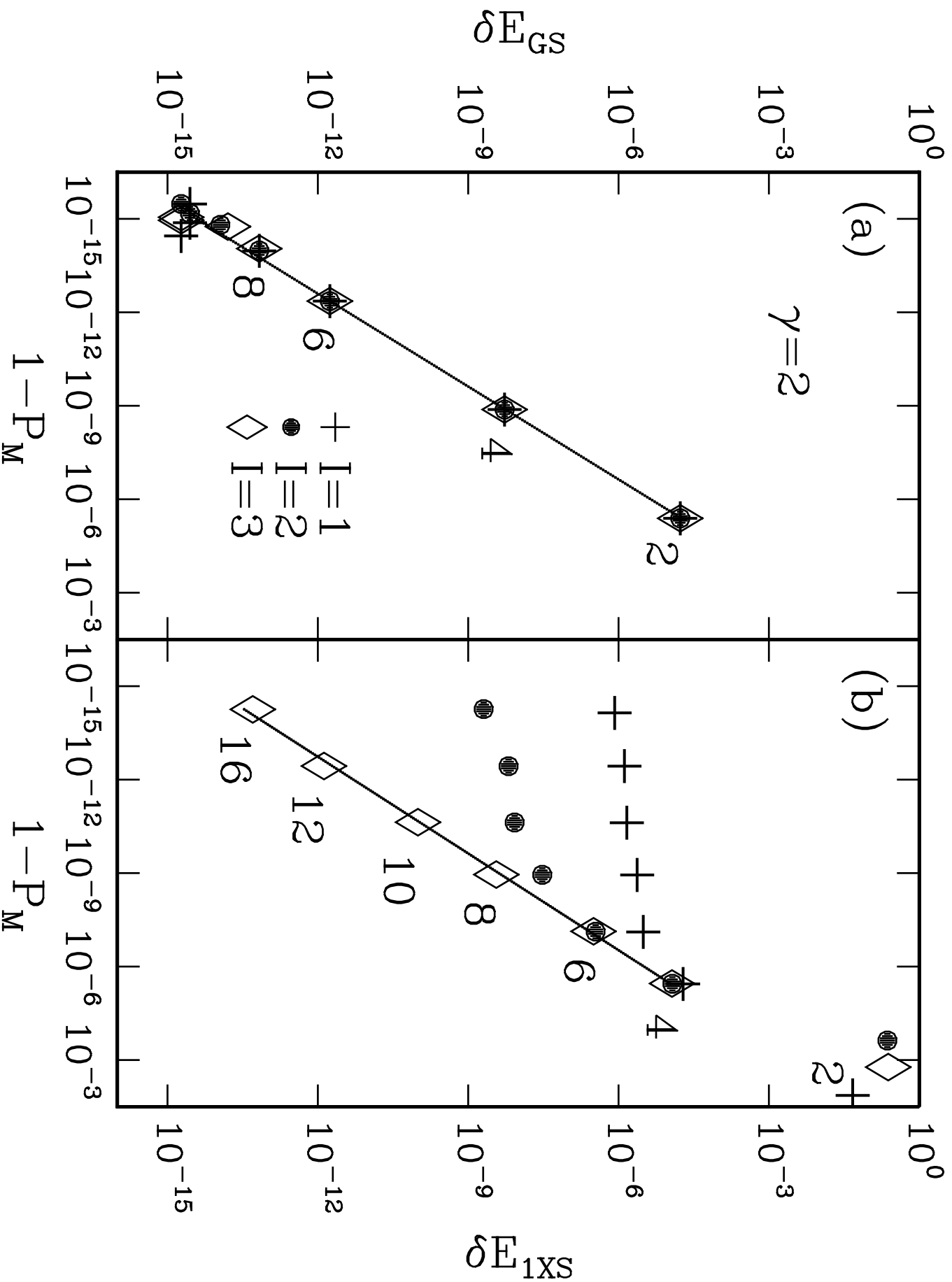
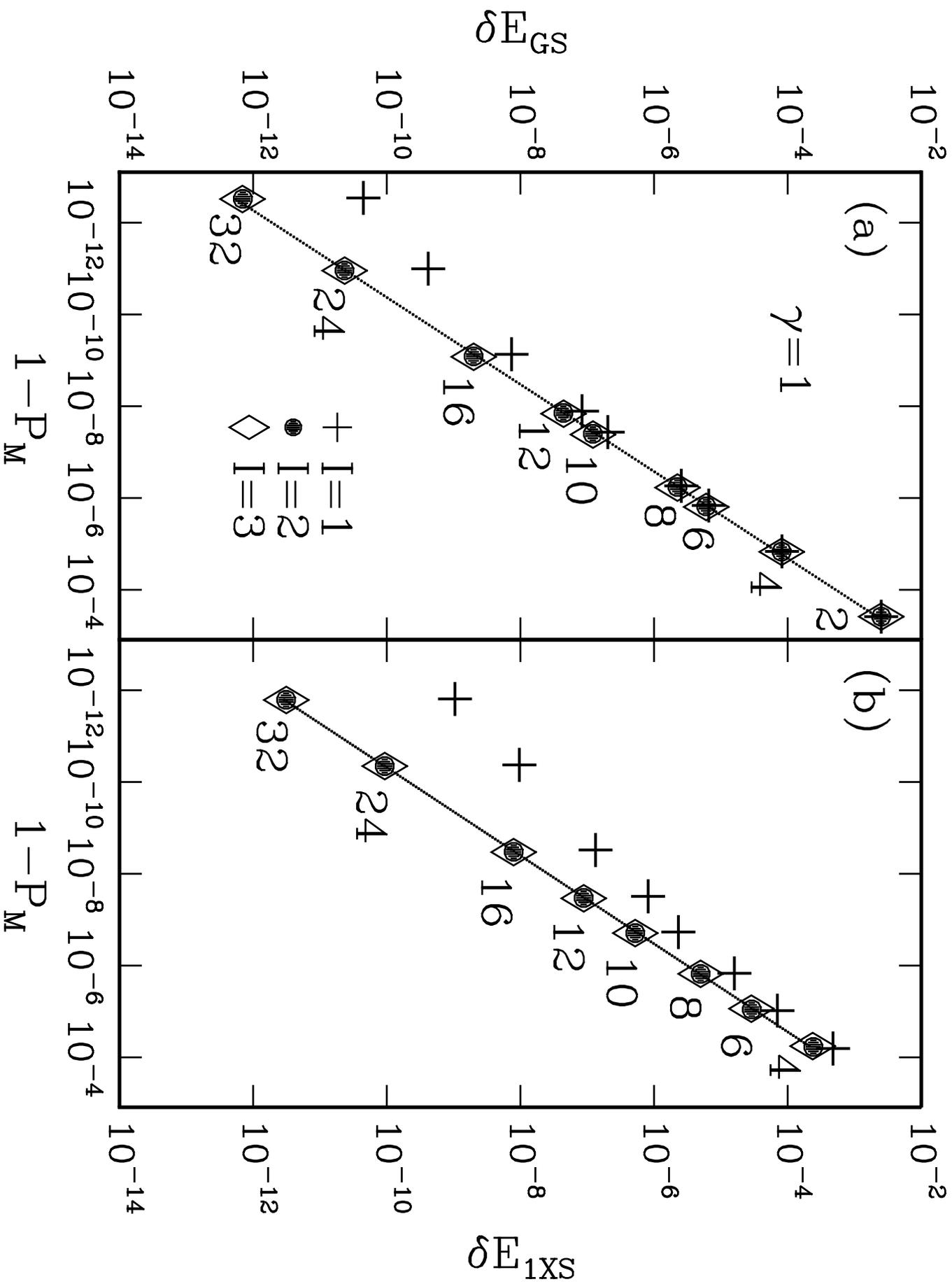
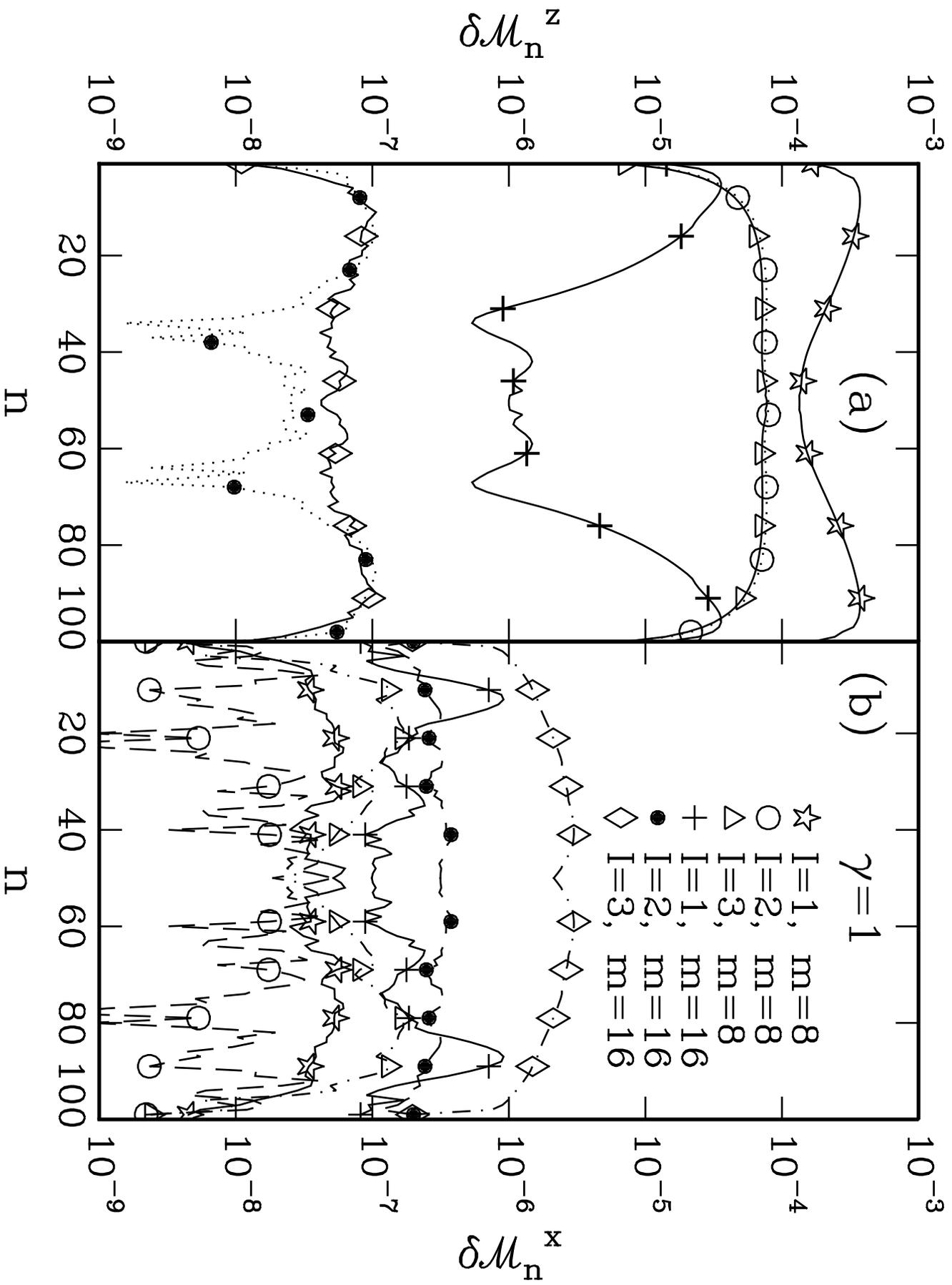
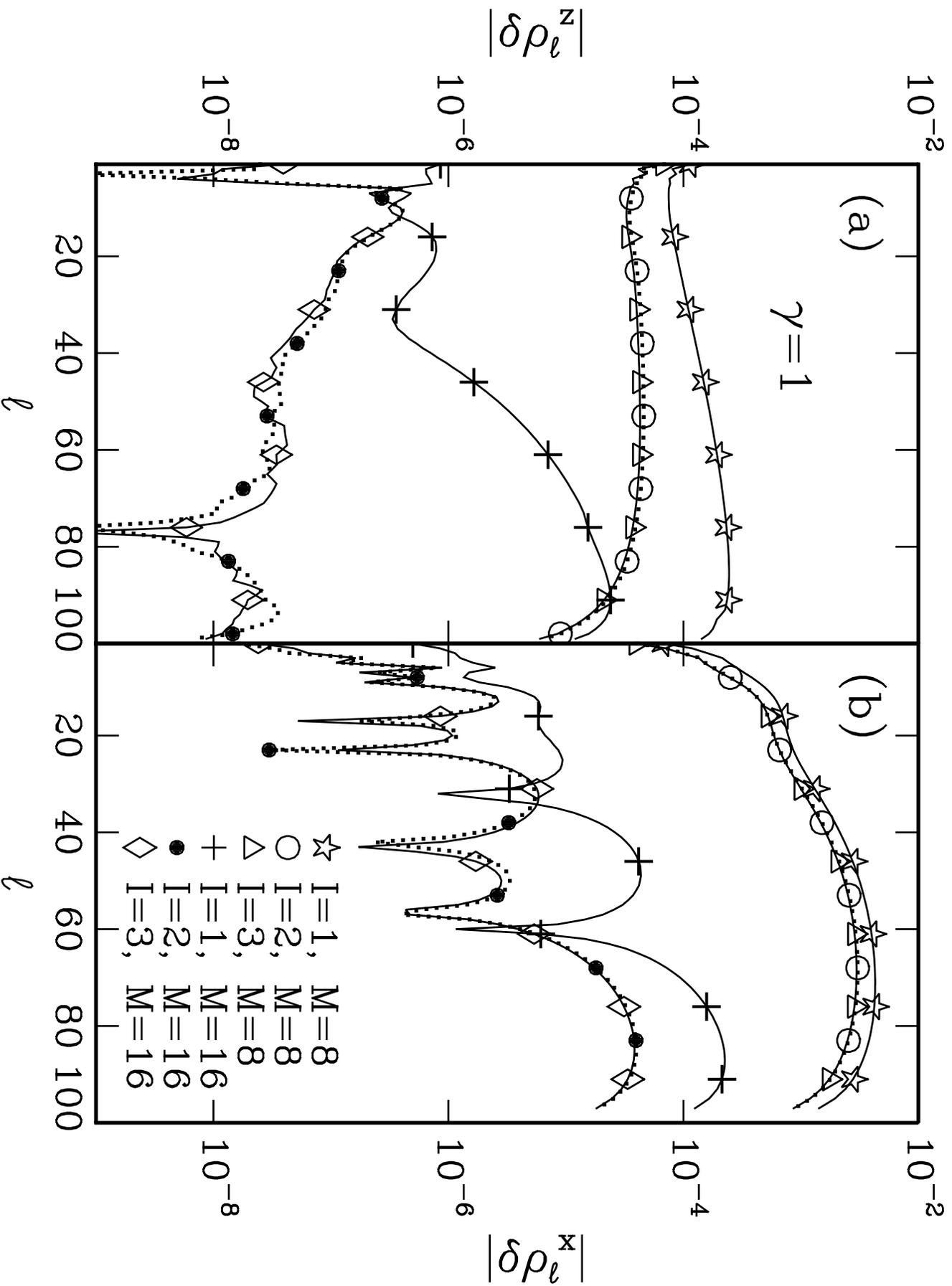
\end{document}